# Shallow EDSLs and Object-Oriented Programming

## Beyond Simple Compositionality

## Weixin Zhang[a] and Bruno C. d. S. Oliveira[a]

a   The University of Hong Kong, Hong Kong, China

**Abstract**   **Context.** Embedded Domain-Specific Languages (EDSLs) are a common and widely used approach to DSLs in various languages, including Haskell and Scala. There are two main implementation techniques for EDSLs: *shallow embeddings* and *deep embeddings*.

**Inquiry.** Shallow embeddings are quite simple, but they have been criticized in the past for being quite limited in terms of modularity and reuse. In particular, it is often argued that supporting multiple DSL interpretations in shallow embeddings is difficult.

**Approach.** This paper argues that shallow EDSLs and Object-Oriented Programming (OOP) are closely related. Gibbons and Wu already discussed the relationship between shallow EDSLs and procedural abstraction, while Cook discussed the connection between procedural abstraction and OOP. We make the transitive step in this paper by connecting shallow EDSLs directly to OOP via procedural abstraction. The knowledge about this relationship enables us to improve on implementation techniques for EDSLs.

**Knowledge.** This paper argues that common OOP mechanisms (including *inheritance*, *subtyping*, and *type-refinement*) increase the modularity and reuse of shallow EDSLs when compared to classical procedural abstraction by enabling a simple way to express *multiple, possibly dependent, interpretations*.

**Grounding.** We make our arguments by using Gibbons and Wu's examples, where procedural abstraction is used in Haskell to model a simple shallow EDSL. We recode that EDSL in Scala and with an improved OO-inspired Haskell encoding. We further illustrate our approach with a case study on refactoring a deep external SQL query processor to make it more modular, shallow, and embedded.

**Importance.** This work is important for two reasons. Firstly, from an intellectual point of view, this work establishes the connection between shallow embeddings and OOP, which enables a better understanding of both concepts. Secondly, this work illustrates programming techniques that can be used to improve the modularity and reuse of shallow EDSLs.

# The Art, Science, and Engineering of Programming

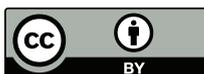

## 1 Introduction

Since Hudak's seminal paper [10] on embedded domain-specific languages (EDSLs), existing languages have been used to directly encode DSLs. Two common approaches to EDSLs are the so-called *shallow* and *deep* embeddings. Deep embeddings emphasize a *syntax*-first approach: the abstract syntax is defined first using a data type, and then interpretations of the abstract syntax follow. The role of interpretations in deep embeddings is to map syntactic values into semantic values in a semantic domain. Shallow embeddings emphasize a *semantics*-first approach, where a semantic domain is defined first. In the shallow approach, the operations of the EDSLs are interpreted directly into the semantic domain. Therefore there is no data type representing uninterpreted abstract syntax.

The trade-offs between shallow and deep embeddings have been widely discussed [20, 11]. Deep embeddings enable transformations on the abstract syntax tree (AST), and multiple interpretations are easy to implement. Shallow embeddings enforce the property of *compositionality* by construction and are easily extended with new EDSL operations. Such discussions lead to a generally accepted belief that it is hard to support multiple interpretations [20] and AST transformations in shallow embeddings.

Compositionality is considered a sign of good language design, and it is one of the hallmarks of denotational semantics. Compositionality means that a denotation (or interpretation) of a language is constructed from the denotation of its parts. Compositionality leads to a modular semantics, where adding new language constructs does not require changes in the semantics of existing constructs. Because compositionality offers a guideline for good language design, Erwig and Walkingshaw [6] argue that a semantics-first approach to EDSLs is superior to a syntax-first approach. Shallow embeddings fit well with such a semantics-driven approach. Nevertheless, the limitations of shallow embeddings compared to deep embeddings can deter their use.

This programming pearl shows that, given adequate language support, having multiple modular interpretations in shallow DSLs is not only possible but simple. Therefore we aim to debunk the belief that multiple interpretations are hard to model with shallow embeddings. Several previous authors [7, 6] already observed that, by using products and projections, multiple interpretations can be supported with a cumbersome and often non-modular encoding. Moreover, it is also known that multiple interpretations *without dependencies* on other interpretations are modularized easily using variants Church encodings [7, 2, 15]. We show that a solution for multiple interpretations, including dependencies, is encodable naturally when the host language combines functional features with common OO features, such as *subtyping*, *inheritance*, and *type-refinement*.

At the center of this pearl is Reynolds' [17] idea of *procedural abstraction*, which enables us to relate shallow embeddings and OOP directly. With procedural abstraction, data is characterized by the operations that are performed over it. This pearl builds on two independently observed connections to procedural abstraction:

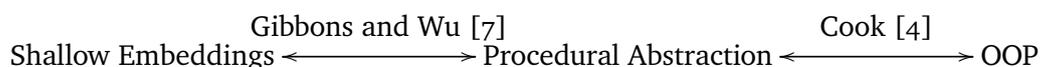

The first connection is between procedural abstraction and shallow embeddings. As Gibbons and Wu [7] state, *"it was probably known to Reynolds, who contrasted deep embeddings ('user defined types') and shallow ('procedural data structures')".* Gibbons and Wu noted the connection between shallow embeddings and procedural abstractions, although they did not go into much detail. The second connection is the connection between OOP and procedural abstraction, which was discussed in depth by Cook [4].

We make our arguments concrete using Gibbons and Wu [7]'s examples, where procedural abstraction is used in Haskell to model a simple *shallow* EDSL. We recode that EDSL in Scala using Wang and Oliveira's [23] extensible interpreter pattern, which provides a simple solution to the *Expression Problem* [22]. The resulting Scala version has *modularity* advantages over the Haskell version, due to the use of subtyping, inheritance, and type-refinement. In particular, the Scala code can easily express modular interpretations that may *not only depend on themselves but also depend on other modular interpretations*, leading to our motto: *beyond simple compositionality*.

While Haskell does not natively support subtyping, inheritance, and type-refinement, its powerful and expressive type system is sufficient to encode similar features. Therefore we can port back to Haskell some of the ideas used in the Scala solution using an improved Haskell encoding that has similar (and sometimes even better) benefits in terms of modularity. In essence, in the Haskell solution we encode a form of subtyping on pairs using type classes. This is useful to avoid explicit projections, that clutter the original Haskell solution. Inheritance is encoded by explicitly delegating interpretations using Haskell superclasses. Finally, type-refinement is simulated using the subtyping typeclass to introduce subtyping constraints.

While the techniques are still cumbersome for transformations, yielding efficient shallow EDSLs is still possible via staging [19, 2]. By removing the limitation of multiple interpretations, we enlarge the applicability of shallow embeddings. A concrete example is our case study, which refactors an external SQL query processor that employs deep embedding techniques [18] into a shallow EDSL. The refactored implementation allows both new (possibly dependent) interpretations and new constructs to be introduced modularly without sacrificing performance. The complete code for all examples and case study is available at https://github.com/wxzh/shallow-dsl.

## 2   Shallow object-oriented programming

This section shows how OOP and shallow embeddings are related via procedural abstraction. We use the same DSL presented by Gibbons and Wu [7] as a running example. We first give the original shallow embedded implementation in Haskell, and rewrite it towards an "OOP style". Then translating the program into a functional OOP language like Scala becomes straightforward.

### 2.1   SCANS: A DSL for parallel prefix circuits

SCANS [8] is a DSL for describing parallel prefix circuits. Given an associative binary operator $\bullet$, the prefix sum of a non-empty sequence $x_1, x_2, ..., x_n$ is $x_1, x_1 \bullet x_2, ..., x_1 \bullet$

⟨*circuit*⟩ ::= 'id' ⟨*positive-number*⟩
      | 'fan' ⟨*positive-number*⟩
      | ⟨*circuit*⟩ 'beside' ⟨*circuit*⟩
      | ⟨*circuit*⟩ 'above' ⟨*circuit*⟩
      | 'stretch' ⟨*positive-numbers*⟩ ⟨*circuit*⟩
      | '(' ⟨*circuit*⟩ ')'

■ **Figure 1** The grammar of Scans.

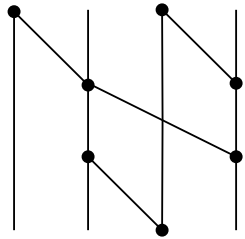

```
(fan 2 beside fan 2) above
(stretch 2 2 fan 2) above
(id 1 beside fan 2 beside id 1)
```

■ **Figure 2** The Brent-Kung circuit of width 4.

$x_2 \bullet ... \bullet x_n$. Such computation can be performed in parallel for a parallel prefix circuit. Parallel prefix circuits have many applications, including binary addition and sorting algorithms. The grammar of Scans is given in Figure 1. Scans has five constructs: two primitives (*id* and *fan*) and three combinators (*beside*, *above* and *stretch*). Their meanings are: *id n* contains $n$ parallel wires; *fan n* has $n$ parallel wires with the leftmost wire connected to all other wires from top to bottom; $c_1$ *beside* $c_2$ joins two circuits $c_1$ and $c_2$ horizontally; $c_1$ *above* $c_2$ combines two circuits of the same width vertically; *stretch ns c* inserts wires into the circuit $c$ so that the $i^{th}$ wire of $c$ is stretched to a position of $ns_1 + ... + ns_i$, resulting in a new circuit of width by summing up *ns*. Figure 2 visualizes a circuit constructed using all these five constructs. The structure of this circuit is explained as follows. The whole circuit is vertically composed by three sub-circuits: the top sub-circuit is a two 2-*fans* put side by side; the middle sub-circuit is a 2-*fan* stretched by inserting a wire on the left-hand side of its first and second wire; the bottom sub-circuit is a 2-*fan* in the middle of two 1-*ids*.

## 2.2 Shallow embeddings and OOP

Shallow embeddings define a language directly by encoding its semantics using procedural abstraction. In the case of Scans, a shallow embedded implementation (in Haskell) conforms to the following types:

```
type Circuit = ...   – the operations we wish to support for circuits
id        :: Int → Circuit
fan       :: Int → Circuit
beside    :: Circuit → Circuit → Circuit
above     :: Circuit → Circuit → Circuit
stretch   :: [Int] → Circuit → Circuit
```

The type *Circuit*, representing the semantic domain, is to be filled with a concrete type according to the semantics. Each construct is declared as a function that produces a *Circuit*. Suppose that the semantics of SCANS calculates the width of a circuit. The definitions are:

```
type Circuit = Int
id n        = n
fan n       = n
beside c₁ c₂ = c₁ + c₂
above c₁ c₂  = c₁
stretch ns c = sum ns
```

For this interpretation, the Haskell domain is simply *Int*. This means that we will get the width immediately after the construction of a circuit. Note that the *Int* domain for *width* is a degenerate case of procedural abstraction: *Int* can be viewed as a no-argument function. In Haskell, due to laziness, *Int* is a good representation. In a call-by-value language, a no-argument function $() \rightarrow Int$ is more appropriate to deal correctly with potential control-flow language constructs.

Now we are able to construct the circuit in Figure 2 using these definitions:

```
> (fan 2 'beside' fan 2) 'above' stretch [2,2] (fan 2) 'above' (id 1 'beside' fan 2 'beside' id 1)
4
```

**Towards OOP**    An *isomorphic encoding* of *width* is given below, where a record with one field captures the domain and is declared as a **newtype**:

```
newtype Circuit₁ = Circuit₁ {width₁ :: Int}
id₁ n          = Circuit₁ {width₁ = n}
fan₁ n         = Circuit₁ {width₁ = n}
beside₁ c₁ c₂   = Circuit₁ {width₁ = width₁ c₁ + width₁ c₂}
above₁ c₁ c₂    = Circuit₁ {width₁ = width₁ c₁}
stretch₁ ns c   = Circuit₁ {width₁ = sum ns}
```

The implementation is still shallow because Haskell's **newtype** does not add any operational behavior to the program. Hence the two programs are effectively the same. However, having fields makes the program look more like an OO program.

**Porting to Scala**    Indeed, we can easily translate the program from Haskell to Scala, as shown in Figure 3. The idea is to map Haskell's record types into an object interface (modeled as a **trait** in Scala) *Circuit₁*, and Haskell's field declarations become method declarations. Object interfaces make the connection to procedural abstraction clear: data is modeled by the operations that can be performed over it. Each case in the semantic function corresponds to a concrete implementation of *Circuit₁*, where function parameters are captured as immutable fields.

This implementation is essentially how we would model SCANS with an OOP language in the first place. A minor difference is the use of traits instead of classes in implementing *Circuit₁*. Although a class definition like

```
// object interface
trait Circuit₁ {def width : Int}
// concrete implementations
trait Id₁ extends Circuit₁ {
    val n : Int
    def width = n
}
trait Fan₁ extends Circuit₁ {
    val n : Int
    def width = n
}
```

```
trait Beside₁ extends Circuit₁ {
    val c₁, c₂ : Circuit₁
    def width = c₁.width + c₂.width
}
trait Above₁ extends Circuit₁ {
    val c₁, c₂ : Circuit₁
    def width = c₁.width
}
trait Stretch₁ extends Circuit₁ {
    val ns : List[Int]; val c : Circuit₁
    def width = ns.sum
}
```

■ **Figure 3** Circuit interpretation in Scala.

```
class Id₁(n : Int) extends Circuit₁ {def width = n}
```

is more common, some modularity offered by the trait version (e.g. mixin composition) is lost. To use this Scala implementation in a manner similar to the Haskell implementation, we need some smart constructors for creating objects conveniently:

```
def id(x : Int)                          = new Id₁     {val n = x}
def fan(x : Int)                         = new Fan₁    {val n = x}
def beside(x : Circuit₁, y : Circuit₁)   = new Beside₁ {val c₁ = x; val c₂ = y}
def above(x : Circuit₁, y : Circuit₁)    = new Above₁  {val c₁ = x; val c₂ = y}
def stretch(x : Circuit₁, xs : Int*)     = new Stretch₁ {val ns = xs.toList; val c = x}
```

Now we are able to construct the circuit shown in Figure 2 in Scala:

```
val circuit = above(beside(fan(2), fan(2)),
                    above(stretch(fan(2), 2, 2),
                        beside(beside(id(1), fan(2)), id(1))))
```

Finally, calling *circuit.width* will return 4 as expected.

As this example illustrates, shallow embeddings and straightforward OO programming are closely related. The syntax of the Scala code is not as concise as the Haskell version due to some extra verbosity caused by trait declarations and smart constructors. Nevertheless, the code is still quite compact and elegant, and the Scala implementation has advantages in terms of modularity, as we shall see next.

## 3 Multiple interpretations in shallow embeddings

An often stated limitation of shallow embeddings is that multiple interpretations are difficult. Gibbons and Wu [7] work around this problem by using tuples. However, their encoding needs to modify the original code and thus is non-modular. This section illustrates how various types of interpretations can be *modularly* defined using standard OOP mechanisms, and compares the result with Gibbons and Wu's Haskell implementations.

### 3.1 Simple multiple interpretations

A single interpretation may not be enough for realistic DSLs. For example, besides *width*, we may want to have another interpretation that calculates the depth of a circuit in SCANS.

**Multiple interpretations in Haskell**   Here is Gibbons and Wu [7]'s solution:

$$\textbf{type } Circuit_2 = (Int, Int)$$
$$id_2\ n \qquad = (n, 0)$$
$$fan_2\ n \qquad = (n, 1)$$
$$above_2\ c_1\ c_2 = (width\ c_1, depth\ c_1 + depth\ c_2)$$
$$beside_2\ c_1\ c_2 = (width\ c_1 + width\ c_2, depth\ c_1\ `max`\ depth\ c_2)$$
$$stretch_2\ ns\ c = (sum\ ns, depth\ c)$$

$$width = fst$$
$$depth = snd$$

A tuple is used to accommodate multiple interpretations, and each interpretation is defined as a projection on the tuple. However, this solution is not modular because it relies on defining the two interpretations (*width* and *depth*) simultaneously. It is not possible to reuse the independently defined *width* interpretation in Section 2.2. Whenever a new interpretation is needed (e.g. *depth*), the original code has to be revised: the arity of the tuple must be incremented and the new interpretation has to be appended to each case.

**Multiple interpretations in Scala**   In contrast, a Scala solution allows new interpretations to be introduced in a modular way:

```scala
trait Circuit₂ extends Circuit₁ {def depth : Int} // subtyping
trait Id₂ extends Id₁ with Circuit₂ {def depth = 0}
trait Fan₂ extends Fan₁ with Circuit₂ {def depth = 1}
trait Above₂ extends Above₁ with Circuit₂ {      // inheritance
    override val c₁, c₂ : Circuit₂                    // covariant type-refinement
    def depth = c₁.depth + c₂.depth
}
trait Beside₂ extends Beside₁ with Circuit₂ {
    override val c₁, c₂ : Circuit₂
    def depth = Math.max (c₁.depth, c₂.depth)
}
trait Stretch₂ extends Stretch₁ with Circuit₂ {
    override val c : Circuit₂
```

    **def** $depth = c.depth$
}

The encoding relies on three OOP abstraction mechanisms: *inheritance*, *subtyping*, and *type-refinement*. Specifically, $Circuit_2$ is a subtype of $Circuit_1$, which extends the semantic domain with a *depth* method. Concrete cases, for instance $Above_2$, implement $Circuit_2$ by inheriting $Above_1$ and implementing *depth*. Also, fields of type $Circuit_1$ are covariantly refined as type $Circuit_2$ to allow *depth* invocations. Importantly, all definitions for *width* in Section 2.2 are *modularly reused* here.

### 3.2 Dependent interpretations

*Dependent interpretations* are a generalization of multiple interpretations. A dependent interpretation does not only depend on itself but also on other interpretations, which goes beyond simple compositional interpretations. An instance of dependent interpretation is *wellSized*, which checks whether a circuit is constructed correctly. The interpretation of *wellSized* is dependent because combinators like *above* use *width* in their definitions.

**Dependent interpretations in Haskell**   In Gibbons and Wu Haskell's solution, dependent interpretations are again defined with tuples in a non-modular way:

**type** $Circuit_3 = (Int, Bool)$
$id_3\ n$          $= (n, True)$
$fan_3\ n$         $= (n, True)$
$above_3\ c_1\ c_2$   $= (width\ c_1, wellSized\ c_1 \wedge wellSized\ c_2 \wedge width\ c_1 \equiv width\ c_2)$
$beside_3\ c_1\ c_2$  $= (width\ c_1 + width\ c_2, wellSized\ c_1 \wedge wellSized\ c_2)$
$stretch_3\ ns\ c$   $= (sum\ ns, wellSized\ c \wedge length\ ns \equiv width\ c)$

$wellSized = snd$

where *width* is called in the definition of *wellSized* for $above_3$ and $stretch_3$.

**Dependent interpretations in Scala**   Once again, it is easy to model dependent interpretation with a simple OO approach:

**trait** $Circuit_3$ **extends** $Circuit_1$ {**def** $wellSized : Boolean$} *// dependency declaration*
**trait** $Id_3$ **extends** $Id_1$ **with** $Circuit_3$ {**def** $wellSized = true$}
**trait** $Fan_3$ **extends** $Fan_1$ **with** $Circuit_3$ {**def** $wellSized = true$}
**trait** $Above_3$ **extends** $Above_1$ **with** $Circuit_3$ {
    **override val** $c_1, c_2 : Circuit_3$
    **def** $wellSized =$
        $c_1.wellSized \wedge c_2.wellSized \wedge c_1.width \equiv c_2.width$     *// dependency usage*
}
**trait** $Beside_3$ **extends** $Beside_1$ **with** $Circuit_3$ {
    **override val** $c_1, c_2 : Circuit_3$
    **def** $wellSized = c_1.wellSized \wedge c_2.wellSized$

```
    }
trait Stretch₃ extends Stretch₁ with Circuit₃ {
    override val c : Circuit₃
    def wellSized = c.wellSized ∧ ns.length ≡ c.width        // dependency usage
}
```

Note that *width* and *wellSized* are defined separately. Essentially, it is sufficient to define *wellSized* while knowing only the signature of *width* in the object interface. In the definition of $Above_3$, for example, it is possible not only to call *wellSized*, but also *width*.

### 3.3 Context-sensitive interpretations

Interpretations may rely on some context. Consider an interpretation that simplifies the representation of a circuit. A circuit can be divided horizontally into layers. Each layer can be represented as a sequence of pairs $(i, j)$, denoting the connection from wire $i$ to wire $j$. For instance, the circuit shown in Figure 2 has the following layout:

$$[[(0,1),(2,3)],[(1,3)],[(1,2)]]$$

The combinator *stretch* and *beside* will change the layout of a circuit. For example, if two circuits are put side by side, all the indices of the right circuit will be increased by the width of the left circuit. Hence the interpretation *layout* is also dependent, relying on itself as well as *width*. An intuitive implementation of *layout* performs these changes immediately to the affected circuit. A more efficient implementation accumulates these changes and applies them all at once. Therefore, an accumulating parameter is used to achieve this goal, which makes *layout* context-sensitive.

**Context-sensitive interpretations in Haskell**    The following Haskell code implements (non-modular) *layout*:

```
type Circuit₄ = (Int, (Int → Int) → [[(Int, Int)]])
id₄ n          = (n, λf → [ ])
fan₄ n         = (n, λf → [[(f 0, f j) | j ← [ 1 .. n − 1 ]]])
above₄ c₁ c₂   = (width c₁, λf → layout c₁ f ++ layout c₂ f)
beside₄ c₁ c₂  = (width c₁ + width c₂,
                  λf → lzw (++) (layout c₁ f) (layout c₂ (f ∘ (width c₁+))))
stretch₄ ns c  = (sum ns, λf → layout c (f ∘ pred ∘ (scanl1 (+) ns!!)))
lzw            :: (a → a → a) → [a] → [a] → [a]
lzw f [ ] ys   = ys
lzw f xs [ ]   = xs
lzw f (x : xs) (y : ys) = f x y : lzw f xs ys

layout = snd
```

The domain of *layout* is a function type $(Int → Int) → [[(Int, Int)]]$, which takes a transformation on wires and produces a layout. An anonymous function is hence defined for each case, where $f$ is the accumulating parameter. Note that $f$ is accumulated in $beside_4$ and $stretch_4$ through function composition, propagated in $above_4$, and finally applied to wire connections in $fan_4$. An auxiliary definition $lzw$ (stands for "long zip with") zips two lists by applying the binary operator to elements of the same index and appending the remaining elements from the longer list to the resulting list. By calling $layout$ on a circuit and supplying an identity function as the initial value of the accumulating parameter, we will get the layout.

**Context-sensitive interpretations in Scala**    Context-sensitive interpretations in Scala are unproblematic as well:

> **trait** $Circuit_4$ **extends** $Circuit_1$ $\{$ **def** $layout(f : Int \Rightarrow Int) : List[List[(Int, Int)]]\}$
> **trait** $Id_4$ **extends** $Id_1$ **with** $Circuit_4$ $\{$ **def** $layout(f : Int \Rightarrow Int) = List()\}$
> **trait** $Fan_4$ **extends** $Fan_1$ **with** $Circuit_4$ $\{$
>     **def** $layout(f : Int \Rightarrow Int) = List($**for**$(i \leftarrow List.range(1, n))$ **yield** $(f(0), f(i)))$
> $\}$
> **trait** $Above_4$ **extends** $Above_1$ **with** $Circuit_4$ $\{$
>     **override val** $c_1, c_2 : Circuit_4$
>     **def** $layout(f : Int \Rightarrow Int) = c_1.layout(f) + c_2.layout(f)$
> $\}$
> **trait** $Beside_4$ **extends** $Beside_1$ **with** $Circuit_4$ $\{$
>     **override val** $c_1, c_2 : Circuit_4$
>     **def** $layout(f : Int \Rightarrow Int) =$
>         $lzw(c_1.layout(f), c_2.layout(f.compose(c_1.width + \_)))(\_ +\!\!+ \_)$
> $\}$
> **trait** $Stretch_4$ **extends** $Stretch_1$ **with** $Circuit_4$ $\{$
>     **override val** $c : Circuit_4$
>     **def** $layout(f : Int \Rightarrow Int) = \{$
>         **val** $vs = ns.scanLeft(0)(\_ + \_).tail$
>         $c.layout(f.compose(vs(\_) - 1))\}$
> $\}$
> **def** $lzw[A](xs : List[A], ys : List[A])(f :(A, A) \Rightarrow A) : List[A] = (xs, ys)$ **match** $\{$
>     **case**$(Nil, \_)$        $\Rightarrow ys$
>     **case**$(\_, Nil)$        $\Rightarrow xs$
>     **case**$(x :: xs, y :: ys) \Rightarrow f(x, y) :: lzw(xs, ys)(f)$
> $\}$

The Scala version captures contexts as method arguments and the implementation of $layout$ is a direct translation from the Haskell version. There are some minor syntax differences that need explanations. Firstly, in $Fan_4$, a **for** *comprehension* is used for producing a list of connections. Secondly, for simplicity, anonymous functions are created without a parameter list. For example, inside $Beside_4$, $c_1.width + \_$ is a shorthand for $i \Rightarrow c_1.width + i$, where the placeholder $\_$ plays the role of the named parameter $i$. Thirdly, function composition is achieved through the *compose* method defined on function values, which has a reverse composition order as opposed to ∘ in Haskell. Fourthly, $lzw$ is implemented as a *curried function*, where the binary operator $f$ is moved to the end as a separate parameter list for facilitating type inference.

### 3.4 An alternative encoding of modular interpretations

There is an alternative encoding of modular interpretations in Scala. For example, the
*wellSized* interpretation can be re-defined like this:

> **trait** $Circuit_3$ **extends** $Circuit_1$ { **def** *wellSized* : *Boolean* }
> **trait** $Id_3$ **extends** $Circuit_3$ { **def** *wellSized* = *true* }
> ...
> **trait** $Stretch_3$ **extends** $Circuit_3$ {
>   **val** $c$ : $Circuit_3$; **val** $ns$ : $List[Int]$
>   **def** *wellSized* = *c.wellSized* $\land$ *ns.length* $\equiv$ *c.width*
> }

where a concrete case like $Id_3$ does not inherit $Id_1$ and leaves the *width* method
unimplemented. Then, an extra step to combine *wellSized* and *width* is needed:

> **trait** $Id_{13}$ **extends** $Id_1$ **with** $Id_3$
> ...
> **trait** $Stretch_{13}$ **extends** $Stretch_1$ **with** $Stretch_3$

Compared to the previous encoding, this encoding is more modular because it
decouples *wellSized* with a particular implementation of *width*. However, more boiler-
plate is needed for combining interpretations. Moreover, it requires some support for
*multiple-inheritance*, which restricts the encoding itself from being applied to a wider
range of OO languages.

### 3.5 Modular language constructs

Besides new interpretations, new language constructs may be needed when a DSL
evolves. For example, in the case of Scans, we may want a *rstretch* (right stretch)
combinator which is similar to the *stretch* combinator but stretches a circuit oppositely.

**New constructs in Haskell**   Shallow embeddings make the addition of *rstretch* easy by
defining a new function:

> *rstretch*    :: $[Int] \rightarrow Circuit_4 \rightarrow Circuit_4$
> *rstretch ns c* = $stretch_4$ $(1 : init\ ns)$ *c* `$beside_4$` $id_4$ $(last\ ns - 1)$

*rstretch* happens to be syntactic sugar over existing constructs. For non-sugar constructs,
a new function that implements all supported interpretations is needed.

**New constructs in Scala**   Such simplicity of adding new constructs is retained in Scala.
Differently from the Haskell approach, there is a clear distinction between syntactic
sugar and ordinary constructs in Scala.

In Scala, syntactic sugar is defined as a smart constructor upon other smart con-
structors:

> **def** $rstretch(ns : List[Int], c : Circuit_4) = stretch(1 :: ns.init, beside(c, id(ns.last - 1)))$

On the other hand, adding ordinary constructs is done by defining a new trait that implements $Circuit_4$. If we treated $rstretch$ as an ordinary construct, its definition would be:

```
trait RStretch extends Stretch₄ {
    override def layout(f : Int ⇒ Int) = {
        val vs = ns.scanLeft(ns.last − 1)(_ + _).init
        c.layout(f.compose(vs(_)))}
}
```

Such an implementation of $RStretch$ illustrates another strength of the Scala approach regarding modularity. Note that $RStretch$ does not implement $Circuit_4$ directly. Instead, it inherits $Stretch_4$ and overrides the $layout$ definition so as to reuse other interpretations as well as field declarations from $Stretch_4$. Inheritance and method overriding enable partial reuse of an existing language construct implementation, which is particularly useful for defining specialized constructs.

### 3.6 Discussion

Gibbons and Wu claim that in shallow embeddings new language constructs are easy to add, but new interpretations are hard. It is possible to define multiple interpretations via tuples, "*but this is still a bit clumsy: it entails revising existing code each time a new interpretation is added, and wide tuples generally lack good language support*" [7]. In other words, Haskell's approach based on tuples is essentially non-modular. However, as our Scala code shows, using OOP mechanisms both language constructs and interpretations are easy to add in shallow embeddings. Moreover, dependent interpretations are possible too, which enables interpretations that may depend on other modular interpretations and go beyond simple compositionality. The key point is that procedural abstraction combined with OOP features (subtyping, inheritance, and type-refinement) adds expressiveness over traditional procedural abstraction.

One worthy point about the Scala solution presented so far is that it is straightforward using OOP mechanisms, it uses only simple types, and dependent interpretations are not a problem. Gibbons and Wu do discuss a number of more advanced techniques [2, 21] that can solve *some* of the modularity problems. In their paper, they show how to support modular *depth* and *width* (corresponding to Section 3.1) using the Finally Tagless [2] approach. This is possible because *depth* and *width* are non-dependent. However they do not show how to modularize *wellSized* nor *layout* (corresponding to Section 3.2 and 3.3, respectively). In Section 4 we revisit such Finally Tagless encoding and improve it to allow dependent interpretations, inspired by the OO solution presented in this section.

## 4   Modular interpretations in Haskell

Modular interpretations are also possible in Haskell via a variant of Church encodings that uses type classes. The original technique is due to Hinze [9] and was shown to be modular and extensible by Oliveira, Hinze and Löh [16]. It has since been popularized under the name Finally Tagless [2] in the context of embedded DSLs. The idea is to use a *type class* to abstract over the signatures of constructs and define interpretations as instances of that type class. This section recodes the SCANS example and compares the two modular implementations in Haskell and Scala.

### 4.1 Revisiting SCANS

Here is the type class defined for SCANS:

```
class Circuit c where
    id     :: Int → c
    fan    :: Int → c
    above  :: c → c → c
    beside :: c → c → c
    stretch :: [Int] → c → c
```

The signatures are the same as what Section 2.2 shows except that the semantic domain is captured by a type parameter $c$. Interpretations such as *width* are then defined as instances of *Circuit*:

```
newtype Width = Width {width :: Int}

instance Circuit Width where
    id n       = Width n
    fan n      = Width n
    above c₁ c₂ = Width (width c₁)
    beside c₁ c₂ = Width (width c₁ + width c₂)
    stretch ns c = Width (sum ns)
```

where $c$ is instantiated as a record type *Width*. Instantiating the type parameter as *Width* rather than *Int* avoids the conflict with the *depth* interpretation which also produces integers.

**Multiple interpretations**   Adding the *depth* interpretation can now be done in a modular manner similar to *width*:

```
newtype Depth = Depth {depth :: Int}

instance Circuit Depth where
    id n        = Depth 0
    fan n       = Depth 1
    above c₁ c₂ = Depth (depth c₁ + depth c₂)
    beside c₁ c₂ = Depth (depth c₁ `max` depth c₂)
    stretch ns c = Depth (depth c)
```

## 4.2 Modular dependent interpretations

Adding a modular dependent interpretation like *wellSized* is more challenging in the Finally Tagless approach. However, inspired by the OO approach we can try to mimic the OO mechanisms in Haskell to obtain similar benefits in Haskell. In what follows we explain how to encode subtyping, inheritance, and type-refinement in Haskell and how that encoding enables additional modularity benefits in Haskell.

**Subtyping**    In the Scala solution subtyping avoids the explicit projections that are needed in the Haskell solution presented in Section 3. We can obtain a similar benefit in Haskell by encoding a subtyping relation on tuples in Haskell. We use the following type class, which was introduced by Bahr and Hvitved [1], to express a subtyping relation on tuples:

> **class** $a \prec b$ **where**
>     $prj :: a \to b$
> **instance** $a \prec a$ **where**
>     $prj\ x = x$
> **instance** $(a, b) \prec a$ **where**
>     $prj = fst$
> **instance** $(b \prec c) \Rightarrow (a, b) \prec c$ **where**
>     $prj = prj \circ snd$

In essence a type $a$ is a subtype of a type $b$ (expressed as $a \prec b$) if $a$ has *the same or more* tuple components as the type $b$. This subtyping relation is closely related to the elaboration interpretation of *intersection types* proposed by Dunfield [5], where Dunfield's merge operator corresponds (via elaboration) to the tuple constructor and projections are implicit and type-driven. The function *prj* simulates up-casting, which converts a value of type $a$ to a value of type $b$. The three overlapping instances define the behavior of the projection function by searching for the type being projected in a compound type.

**Modular** *wellSized* **and encodings of inheritance and type-refinement**    Now, defining *wellSized* modularly becomes possible:

> **newtype** $WellSized = WellSized\ \{wellSized :: Bool\}$
> **instance** $(Circuit\ c, c \prec Width) \Rightarrow Circuit\ (WellSized, c)$ **where**
>     $id\ n\quad\ \ = (WellSized\ True, id\ n)$
>     $fan\ n\quad = (WellSized\ True, fan\ n)$
>     $above\ c_1\ c_2 = (WellSized\ (gwellSized\ c_1 \wedge gwellSized\ c_2 \wedge gwidth\ c_1 \equiv gwidth\ c_2)$
>         $, above\ (prj\ c_1)\ (prj\ c_2))$
>     $beside\ c_1\ c_2 = (WellSized\ (gwellSized\ c_1 \wedge gwellSized\ c_2), beside\ (prj\ c_1)\ (prj\ c_2))$
>     $stretch\ ns\ c = (WellSized\ (gwellSized\ c \wedge length\ ns \equiv gwidth\ c), stretch\ ns\ (prj\ c))$
> $gwidth :: (c \prec Width) \Rightarrow c \to Int$
> $gwidth = width \circ prj$

```
gwellSized :: (c ≺ WellSized) ⇒ c → Bool
gwellSized = wellSized ∘ prj
```

Essentially, dependent interpretations are still defined using tuples. The dependency on *width* is expressed by constraining the type parameter as $c \prec Width$. Such constraint allows us to simulate the type-refinement of fields in the Scala solution. Although the implementation is modular, it requires some boilerplate. The reuse of *width* interpretation is achieved via delegation, where *prj* needs to be called on each subcircuit. Such explicit delegation simulates the inheritance employed in the Scala solution. Also, auxiliary definitions *gwidth* and *gwellSized* are necessary for projecting the desired interpretations from the constrained type parameter.

### 4.3 Modular terms

As new interpretations may be added later, a problem is how to construct the term that can be interpreted by those new interpretations without reconstruction. We show how to do this for the circuit shown in Figure 2:

```
circuit :: Circuit c ⇒ c
circuit = (fan 2 'beside' fan 2) 'above'
          stretch [2, 2] (fan 2) 'above'
          (id 1 'beside' fan 2 'beside' id 1)
```

Here, *circuit* is a generic circuit that is not tied to any interpretation. When interpreting *circuit*, its type must be instantiated:

```
> width (circuit :: Width)
4
> depth (circuit :: Depth)
3
> gwellSized (circuit :: (WellSized, Width))
True
```

At user-site, *circuit* must be annotated with the target semantic domain so that an appropriate type class instance for interpretation can be chosen.

**Syntax extensions**   This solution also allows us to modularly extend [16] Scans with more language constructs such as *rstretch*:

```
class Circuit c ⇒ ExtendedCircuit c where
    rstretch :: [Int] → c → c
```

Existing interpretations can be modularly extended to handle *rstretch*:

```
instance ExtendedCircuit Width where
    rstretch = stretch
```

Existing circuits can also be reused for constructing circuits in extended Scans:

```
circuit₂ :: ExtendedCircuit c ⇒ c
circuit₂ = rstretch [2, 2, 2, 2] circuit
```

■ **Table 1** Language features needed for modular interpretations: Scala vs. Haskell.

| Goal | Scala | Haskell |
|---|---|---|
| Multiple interpretation | Trait & Type-refinement | Type class |
| Interpretation reuse | Inheritance | Delegation |
| Dependency declaration | Subtyping | Tuples & Type constraints |

## 4.4 Comparing modular implementations using Scala and Haskell

Although both the Scala and Haskell solutions are able to model modular dependent interpretations, they use a different set of language features. Table 1 compares the language features needed by Scala and Haskell. The Scala approach relies on built-in features. In particular, subtyping, inheritance (mixin composition) and type-refinement are all built-in. This makes it quite natural to program the solutions in Scala, without even needing any form of parametric polymorphism. In contrast, the Haskell solution does not have such built-in support for OO features. Subtyping and type-refinement need to be encoded/simulated using parametric polymorphism and type classes. Inheritance is simulated by explicit delegations. The Haskell encoding is arguably conceptually more difficult to understand and use, but it is still quite simple. One interesting feature that is supported in Haskell is the ability to encode modular terms. This relies on the fact that the constructors are overloaded. The Scala solution presented so far does not allow such overloading, so code using constructors is tied with a specific interpretation. In the next section we will see a final refinement of the Scala solution that enables modular terms, also by using overloaded constructors.

## 5 Modular terms in Scala

One advantage of the Finally Tagless approach over our Scala approach presented so far is that terms can be constructed modularly without tying those terms to any interpretation. Modular terms are also possible by combining our Scala approach with Object Algebras [15], which employ a technique similar to Finally Tagless in the context of OOP. Differently from the Haskell solution presented in Section 4, the Scala approach only employs parametric polymorphism to overload the constructors. Both inheritance and type-refinement do not need to be simulated or encoded.

**Object Algebra interface**    To capture the generic interface of the constructors we define an abstract factory (or Object Algebra interface) for circuits similar to the type class version shown in Section 4.1:

```scala
trait Circuit[C] {
    def id(x : Int) : C
    def fan(x : Int) : C
    def above(x : C, y : C) : C
    def beside(x : C, y : C) : C
```

    $\mathbf{def}\ stretch\,(x:C, xs:Int*):C$
}

which exposes factory methods for each circuit construct supported by Scans.

**Abstract terms**    Modular terms can be constructed via the abstract factory. For example, the circuit shown in Figure 2 is built as:

$\mathbf{def}\ circuit\,[C]\,(f:Circuit\,[C]) =$
    $f.above\,(f.beside\,(f.fan\,(2), f.fan\,(2)),$
            $f.above\,(f.stretch\,(f.fan\,(2), 2, 2),$
                    $f.beside\,(f.beside\,(f.id\,(1), f.fan\,(2)), f.id\,(1))))$

Similarly, *circuit* is a generic method that takes a *Circuit* instance and builds a circuit through that instance. With Scala the definition of *circuit* can be even simpler: we can avoid prefixing "*f*." everywhere by importing *f*. Nevertheless, the definition shown here is more language-independent.

**Object Algebras**    We need concrete factories (Object Algebras) that implement *Circuit* to actually invoke *circuit*. Here is a concrete factory that produces $Circuit_1$:

    $\mathbf{trait}\ Factory_1\ \mathbf{extends}\ Circuit\,[Circuit_1]\ \{\dots\}$

where the omitted code is identical to the smart constructors presented in Section 2.2. Concrete factories for other circuit implementations can be defined in a similar way by instantiating the type parameter *Circuit* accordingly:

    $\mathbf{trait}\ Factory_4\ \mathbf{extends}\ Circuit\,[Circuit_4]\ \{\dots\}$

**Concrete terms**    By supplying concrete factories to abstract terms, we obtain concrete terms that can be interpreted differently:

    $circuit\,(\mathbf{new}\ Factory_1\ \{\}).width$           //4
    $circuit\,(\mathbf{new}\ Factory_4\ \{\}).layout\ \{x \Rightarrow x\}$//List(List((0,1),(2,3)),List((1,3)),List((1,2)))

**Modular extensions**    Both factories and terms can be *modularly* reused when the DSL is extended with new language constructs. To support right stretch for Scans, we first extend the abstract factory with new factory methods:

    $\mathbf{trait}\ ExtendedCircuit\,[C]\ \mathbf{extends}\ Circuit\,[C]\ \{\mathbf{def}\ rstretch\,(x:C, xs:Int*):C\}$

We can also build extended concrete factories upon existing concrete factories:

    $\mathbf{trait}\ ExtendedFactory_4\ \mathbf{extends}\ ExtendedCircuit\,[Circuit_4]\ \mathbf{with}\ Factory_4\ \{$
        $\mathbf{def}\ rstretch\,(x:Circuit_4, xs:Int*) = \mathbf{new}\ RStretch\ \{\mathbf{val}\ c = x; \mathbf{val}\ ns = xs.toList\}$
    }

Furthermore, previously defined terms can be reused in constructing extended terms:

    $\mathbf{def}\ circuit_2\,[C]\,(f:ExtendedCircuit\,[C]) = f.rstretch\,(circuit\,(f), 2, 2, 2, 2)$

### 6 Case study: a shallow EDSL for SQL queries

A common motivation for using deep embeddings is performance. Deep embeddings enable complex AST transformations, which is useful to implement optimizations that improve the performance. An alternative way to obtain performance is to use staging frameworks, such as Lightweight Modular Staging (LMS) [19]. As illustrated by Rompf and Amin [18] staging can preclude the need for AST transformations for a realistic query DSL. To further illustrate the applicability of shallow OO embeddings, we refactored Rompf and Amin's deep, external DSL implementation to make it more *modular*, *shallow* and *embedded*. The shallow DSL retains the performance of the original deep DSL by generating the same code.

#### 6.1 Overview

SQL is the best-known DSL for data queries. Rompf and Amin [18] present a SQL query processor implementation in Scala. Their implementation is an *external* DSL, which first parses a SQL query into a relational algebra AST and then executes the query in terms of that AST. Based on the LMS framework [19], the SQL compilers are nearly as simple as an interpreter while having performance comparable to hand-written code. The implementation uses deep embedding techniques such as algebraic data types (*case classes* in Scala) and pattern matching for representing and interpreting ASTs. These techniques are a natural choice as multiple interpretations are needed for supporting different backends. But problems arise when the implementation evolves with new language constructs. All existing interpretations have to be modified for dealing with these new cases, suffering from the Expression Problem.

We refactored Rompf and Amin [18]'s implementation into a shallow EDSL for the following reasons. Firstly, multiple interpretations are no longer a problem for our shallow embedding technique. Secondly, the original implementation contains no hand-coded AST transformations. Thirdly, it is common to embed SQL into a general purpose language.

To illustrate our shallow EDSL, suppose there is a data file *talks.csv* that contains a list of talks with time, title and room. We can write several sample queries on this file with our EDSL. A simple query that lists all items in *talks.csv* is:

> **def** $q_0 = FROM$ ("talks.csv")

Another query that finds all talks at 9 am with their room and title selected is:

> **def** $q_1 = q_0$ *WHERE* '*time* === "09:00 AM" *SELECT* ('*room*, '*title*)

Yet another relatively complex query to find all conflicting talks that happen at the same time in the same room with different titles is:

> **def** $q_2 = q_0$ *SELECT* ('*time*, '*room*, '*title AS* '$title_1$)  *JOIN*
> ($q_0$ *SELECT* ('*time*, '*room*, '*title AS* '$title_2$)) *WHERE*
> '$title_1 <>$ '$title_2$

Compared to an external implementation, our embedded implementation has the benefit of reusing the mechanisms provided by the host language for free. As illustrated by the sample queries above, we are able to reuse common subqueries ($q_0$) in building complex queries ($q_1$ and $q_2$). This improves the readability and modularity of the embedded programs.

## 6.2 Embedded syntax

Thanks to the good support for EDSLs in Scala, we can precisely model the syntax of SQL. The syntax of our EDSL is close to that of LINQ [13], where **select** is an optional, terminating clause of a query. We employ well-established OO and Scala techniques to simulate the syntax of SQL queries in our shallow EDSL implementation. Specifically, we use the *Pimp My Library* pattern [14] for lifting field names and literals implicitly. For the syntax of combinators such as **where** and **join**, we adopt a fluent interface style. Fluent interfaces enable writing something like "*FROM*(...).*WHERE*(...).*SELECT*(...)". Scala's infix notation further omits "." in method chains. Other famous embedded SQL implementations in OOP such as LINQ [13] also adopt similar techniques in designing their syntax. The syntax is implemented in a pluggable way, in the sense that the semantics is decoupled from the syntax. Details of the syntax implementation are beyond the scope of this pearl. The interested reader can consult the companion code.

Beneath the surface syntax, a relational algebra operator structure is constructed. For example, we will get the following operator structure for $q_1$:

$$Project(Schema("room", "title"),$$
$$Filter(Eq(Field("time"), Value("09:00 AM")),$$
$$Scan("talks.csv")))$$

## 6.3 A relational algebra compiler

A SQL query can be represented by a relational algebra expression. The basic interface of operators is modeled as follows:

```
trait Operator {
    def resultSchema : Schema
    def execOp(yld : Record ⇒ Unit) : Unit
}
```

Two interpretations, *resultSchema* and *execOp*, need to be implemented for each concrete operator: the former collects a schema for projection; the latter executes actions to the records of the table. Very much like the interpretation *layout* discussed in Section 3.3, *execOp* is both *context-sensitive* and *dependent*: it takes a callback *yld* and accumulates what the operator does to records into *yld* and uses *resultSchema* in displaying execution results. In our implementation *execOp* is indeed introduced as an extension just like *layout*. Here we merge the two interpretations for conciseness of presentation. Some core concrete relational algebra operators are given below:

```
trait Project extends Operator {
    val out, in : Schema; val op : Operator
    def resultSchema = out
    def execOp(yld : Record ⇒ Unit) = op.execOp {rec ⇒ yld(Record(rec(in ), out))}
}
trait Join extends Operator {
    val op₁, op₂ : Operator
    def resultSchema = op₁.resultSchema ++ op₂.resultSchema
    def execOp(yld : Record ⇒ Unit) =
        op₁.execOp {rec₁ ⇒
            op₂.execOp {rec₂ ⇒
                val keys = rec₁.schema intersect rec₂.schema
                if(rec₁(keys) ≡ rec₂(keys))
                    yld(Record(rec₁.fields ++ rec₂.fields, rec₁.schema ++ rec₂.schema))}}
}
trait Filter extends Operator {
    val pred : Predicate; val op : Operator
    def resultSchema = op.resultSchema
    def execOp(yld : Record ⇒ Unit) = op.execOp {rec ⇒ if(pred.eval(rec)) yld(rec)}
}
```

*Project* rearranges the fields of a record; *Join* matches a record against another and combines the two records if their common fields share the same values; *Filter* keeps a record only when it meets a certain predicate. There are also two utility operators, *Print* and *Scan*, for processing inputs and outputs, whose definitions are omitted for space reasons.

**From an interpreter to a compiler**   The query processor presented so far is elegant but unfortunately slow. To achieve better performance, Rompf and Amin extend the SQL processor in various ways. One direction is to turn the slow query interpreter into a fast query compiler by generating specialized low-level code for a given query. With the help of the LMS framework, this task becomes rather easy. LMS provides a type constructor *Rep* for annotating computations that are to be performed in the next stage. The signature of the staged *execOp* is:

$$\textbf{def } execOp(yld : Record \Rightarrow Rep[Unit]) : Rep[Unit]$$

where *Unit* is lifted as $Rep[Unit]$ for delaying the actions on records to the generated code. Two staged versions of *execOp* are introduced for generating Scala and C code respectively. By using the technique presented in Section 3, they are added modularly with existing interpretations such as *resultSchema* reused. The implementation of staged *execOp* is similar to the unstaged counterpart except for minor API differences between staged and unstaged types. Hence the simplicity of the implementation remains. At the same time, dramatic speedups are obtained by switching from interpretation to compilation.

■ **Table 2** SLOC for original (Deep) and refactored (Shallow) versions.

| Source | Functionality | Deep | Shallow |
|---|---|---|---|
| *query_unstaged* | SQL interpreter | 83 | 98 |
| *query_staged* | SQL to Scala compiler | 179 | 194 |
| *query_optc* | SQL to C compiler | 245 | 262 |

**Language extensions**  Rompf and Amin also extend the query processor with two new language constructs, hash joins and aggregates. The introduction of these constructs is done in a modular manner with our approach:

```
trait Group extends Operator {
    val keys, agg : Schema; val op : Operator
    def resultSchema = keys ++ agg
    def execOp (yld : Record ⇒ Unit) {...}
}
trait HashJoin extends Join {
    override def execOp (yld : Record ⇒ Unit) = {
        val keys = op₁.resultSchema intersect op₂.resultSchema
        val hm = new HashMapBuffer (keys, op₁.resultSchema)
        op₁.execOp {rec₁ ⇒
            hm(rec₁(keys)) += rec₁.fields}
        op₂.execOp {rec₂ ⇒
            hm(rec₂(keys)) foreach {rec₁ ⇒
                yld(Record(rec₁.fields ++ rec₂.fields, rec₁.schema ++ rec₂.schema))}}}
}
```

*Group* supports SQL's **group by** clause, which partitions records and sums up specified fields from the composed operator. *HashJoin* is a replacement for *Join*, which uses a hash-based implementation instead of naive nested loops. With inheritance and method overriding, we are able to reuse the field declarations and other interpretations from *Join*.

### 6.4 Evaluation

We evaluate our refactored shallow implementation with respect to the original deep implementation. Both implementations of the DSL (the original and our refactored version) *generate the same code*: thus the performance of the two implementations is similar. Hence we compare the two implementations only in terms of the source lines of code (SLOC). We exclude the code related to surface syntax for the fairness of comparison because our refactored version uses embedded syntax whereas the original uses a parser. As seen in Table 2, our shallow approach takes a dozen more lines of code than the original deep approach for each version of SQL processor. The SLOC expansion is attributed to the fact that functional decomposition (case classes) is more compact than object-oriented decomposition in Scala. Nevertheless, our shallow approach makes it easier to add new language constructs.

## 7 Conclusion

This programming pearl reveals the close correspondence between OOP and shallow embeddings: the essence of both is procedural abstraction. It also showed how OOP increases the modularity of shallow EDSLs. OOP abstractions, including subtyping, inheritance, and type-refinement, bring extra modularity to traditional procedural abstraction. As a result, multiple interpretations are allowed to co-exist in shallow embeddings. Moreover, the multiple interpretations can be *dependent*: an interpretation can depend not only on itself but also on other modular interpretations. Thus the approach presented here allows us to go *beyond simple compositionality*, where interpretations can only depend on themselves.

It has always been a hard choice between shallow and deep embeddings when designing an EDSL: there are some tradeoffs between the two styles. Deep embeddings trade some simplicity and the ability to add new language constructs for some extra power. This extra power enables multiple interpretations, as well as complex AST transformations. As this pearl shows, in languages with OOP mechanisms, multiple (possibly dependent) interpretations are still easy to do with shallow embeddings and the full benefits of an extended form of compositionality still apply. Therefore the motivation to employ deep embeddings becomes weaker than before and mostly reduced to the need for AST transformations. Prior work on the Finally Tagless [12] and Object Algebras [24] approaches already show that AST transformations are still possible in those styles. However this requires some extra machinery, and the line between shallow and deep embeddings becomes quite blurry at that point.

Finally, this work shows a combination of two previously studied solutions to the Expression Problem in OO: the extensible interpreter pattern proposed by Wang and Oliveira [23] and Object Algebras [15]. The combination exploits the advantages of each of the approaches to overcome the limitations of each approach individually. In the original approach by Wang and Oliveira modular terms are hard to model, whereas with Object Algebras a difficulty is modeling modular dependent operations. A closely related technique is employed by Cazzola and Vacchi [3], although in the context of external DSLs. Their technique is slightly different with respect to the extensible interpreter pattern. Essentially while our approach is purely based on subtyping and type-refinement, they use generic types instead to simulate the type-refinement. While the focus of our work is embedded DSLs, the techniques discussed here are useful for other applications, including external DSLs as Cazzola and Vacchi show.

**Acknowledgements** We thank Willam R. Cook, Jeremy Gibbons, Ralf Hinze, Martin Erwig, and the anonymous reviewers of GPCE, ICFP, JFP, and Programming for their valuable comments that significantly improved this work. This work is based on the first author's master thesis [25] and has been funded by Hong Kong Research Grant Council projects number 17210617 and 17258816.

**About the authors**

**Weixin Zhang** is a PhD candidate at the University of Hong Kong. His current research interests are centered around modularity, domain-specific languages and generative programming. You can contact him at wxzhang2@cs.hku.hk and find further information at https://wxzh.github.io.

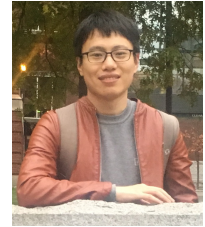

**Bruno C. d. S. Oliveira** is an assistant professor of the at the University of Hong Kong. His research interests are centered around programming languages. His existing research is mainly focused on type systems for modularity and the combination of Object-Oriented and Functional Paradigms. You can contact him at bruno@cs.hku.hk and find further information at https://i.cs.hku.hk/~bruno.

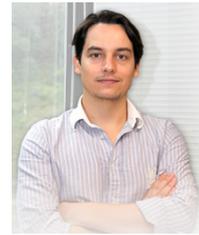